# WIRELESS MESH NETWORK PERFORMANCE FOR URBAN SEARCH AND RESCUE MISSIONS

Cristina Ribeiro[1] Alexander Ferworn[2] and Jimmy Tran[2]

[1]Department of Computer and Information Science, University of Guelph, Guelph, Canada
`cribeiro@uoguelph.ca`
[2]Department of Computer Science, Ryerson University, Toronto, Canada
`aferworn@ryerson.ca, q2tran@ryerson.ca`

**ABSTRACT**

*In this paper we demonstrate that the Canine Pose Estimation (CPE) system can provide a reliable estimate for some poses and when coupled with effective wireless transmission over a mesh network. Pose estimates are time sensitive, thus it is important that pose data arrives at its destination quickly. Propagation delay and packet delivery ratio measuring algorithms were developed and used to appraise Wireless Mesh Network (WMN) performance as a means of carriage for this time-critical data. The experiments were conducted in the rooms of a building where the radio characteristics closely resembled those of a partially collapsed building—a typical US&R environment. This paper presents the results of the experiments, which demonstrate that it is possible to receive the canine pose estimation data in real-time although accuracy of the results depend on the network size and the deployment environment.*

**KEYWORDS**

*Wireless Mesh Network, WiFi, Transmission Delay, Propagation Delay, Packet Delivery Ratio, Wireless Networks for Computational Public Safety, Canine Pose Estimation, Canine Augmentation Technology & Urban Search and Rescue*

## 1. INTRODUCTION

The fastest and most reliable means of finding people trapped after a building collapse is through the use of trained Urban Search and Rescue (USAR) dogs. Sometimes called disaster dogs, these canines are the state-of-the-art when conducting search operations within an urban disaster like those that occurred in Mexico [1], Kobe [2], Turkey [3] or New York [4].

Search operations necessarily occur before rescue can take place. Since there is a finite time that someone can survive entombed within the wreckage of a building, it is critical that search operations occur as quickly and efficiently as possible so that the ensuing operation is rescue and not recovery. Search operations have several challenges that increase the time it takes to find survivors (often called "patients") within the wreckage.

A particular matter requiring improvement is in the situational awareness [5-7] canine handlers have while conducting searches under certain conditions. Situations can arise where a handler is not aware of their dog's whereabouts or behaviour. This lack of situational awareness is generally due to the distance and obstacles between the handler and the canine. In the extreme, a handler may be asked to send his or her dog into the rubble of a building without the ability to actually follow behind, because human access may be precluded or limited. If the handler's situational awareness of the canine could be enhanced, search times could be reduced, improving the performance of the team, resulting in more lives saved.

A complementary area of research is the augmentation of USAR dogs [8-11] with technology that allows emergency first responders to experience what is happening around the dog while it is searching. While this area of research is very important, the handler still does not know what the dog is actually doing while out of sight; this augmenting of senses focuses on what is around the dog and not the dog itself. This technology does however provide some additional situational awareness capabilities, but only from the perspective of the canine.

The orientation of the dog is very important for the handler as the dog's posture communicates a significant amount of information. Orientation or posture, referred to as 'pose', is important, because USAR canines are trained to display different poses to indicate various situational conditions they have experienced. In a sense, they use pose as a language. An example of this is a canine, which is cross-trained to search for cadavers. This specially trained dog assumes the *sitting* pose when it has found a cadaver. Another pose, *lying down*, indicates that the canine has stopped searching because of disinterest, exhaustion, or injury.

Past research has been conducted on animals in terms of behavioural assessments [12]; however, not in the area of situational awareness which is needed for USAR operations. Handlers are limited in their capabilities to conduct searches in cases where their dogs cannot be seen. At the moment there are no solutions that provide the canine handler with situational awareness regarding canine pose.

This paper begins with an overview of Computational Public Safety. In section 2 is a summary of research in the area relating to USAR, and the performance challenges of Wireless Mesh Networks. Section 3 we cover WMN performance experiments conducted in a disaster like environment, not through simulation. We measure and assess performance based on propagation delay, packet delivery ratios and network coverage. A discussion and evaluation of the experimental results are covered in section 4. Section 5 lists the conclusions for this work.

## 1.1 Computational Public Safety

Computational Public Safety (CPS) involves the application of computational resources, theory and practice in support of and improvement to public safety processes. The objective of this work was to develop a new capability ─ to acquire situational awareness in search operations through the determination of canine pose. The work can improve how USAR is conducted by utilizing technology to provide situational awareness to USAR canine handlers, supporting emergency first responders and search managers.

A number of challenges exist in determining canine pose and communicating the relevant information back to the handler. These challenges are: 1) determining canine pose; 2) evaluating the accuracy of the canine pose estimation technique; and 3) evaluating the network's competency to transmit the canine pose data in a timely manner to all essential parties.

A device to read the accelerometer data and transmit the information back to a laptop was designed and constructed. The Canine Pose Estimation (CPE) device is a microcontroller-based device programmed with an algorithm to interpret the raw accelerometer sensor readings. A Wi-Fi device was attached to the microcontroller that enabled the transmission of the canine pose data wirelessly back to a laptop.

A wireless mesh network (WMN) was used to transmit the acceleration data and was comprised of ruggedized mesh routers dispersed around the search area. It was also important to assess the effectiveness of the transmission of the canine pose data across the network. Any significant delays in the transmission would result in the handler not being aware of the canine's behaviour at that particular moment. The repercussions of this would be that the handler would be delayed in reacting to the situation.

To assess the effectiveness of the pose data transmission, a propagation delay measuring algorithm was developed to quantitatively measure the delay across the WMN under different disaster-like conditions. Moreover, the packet delivery ratio (PDR) of the mesh network was determined to further interpret the successful use of this system in disaster situations. By analyzing the PDR we can approximate how much data loss could occur in a real disaster situation and how this may impact the accuracy of the CPE system to determine canine pose.

Finally, a laptop hosted an application with the CPE algorithm. This algorithm took in the raw data and calculated the accelerations, body angles and other variables to determine the dog's pose.

## 2. BACKGROUND

Imagine an earthquake, tsunami, typhoon, hurricane, levy failure, or terrorist bomb attack in an urban centre leaving crumbled walls of concrete, mangled steel frames, and debris blanketing the area. In urban disaster situations such as this, many people may be injured and/or trapped under the rubble unable to escape. In this type of situation, fire fighters, police, emergency medical services and other emergency first responders must work tirelessly around the clock to find and save as many people from the wreckage as possible.

In the first few hours after a disaster, casualties with life-threatening injuries require immediate medical attention if they are to survive. Those who are buried beneath the rubble might survive several weeks without food but only several days without water [13]; more time spent searching for survivors means less time for successful rescue to take place.

### 2.1 Disaster Dogs

Dogs have been used for centuries to assist humans. They have been successfully used for guarding, aiding the blind and hearing impaired, forensic tracking, and for detection of explosives, landmines, narcotics, insect infestations, microbial growth, epilepsy and even cancer detection [14-17]. Their trainability is one of the reasons they can successfully carry out these tasks. Recently, dogs have been augmented with technology to try and provide information regarding their whereabouts to their handler. For example, the United Kingdom Police use FIDO, a camera system, for surveillance in a weapons seizures [18]. This system enabled the handlers to wirelessly monitor the whereabouts of the dog in dangerous circumstances.

In USAR, canines have demonstrated their effectiveness in searching quickly and efficiently [19, 15] and are essentially the search tool of choice. Dogs assets include their highly sensitive sense of smell, enabling them to locate casualties [16-18] and their speed and agility that greatly surpasses that of a human [19]. While these assets are the reason for their effectiveness in searching, their speed and agility also act as impediments in the search process. Often the handler will fall far behind the dog as more than 70 percent of searches are conducted with the dogs off leash [20]. When the dog is out of the line-of-sight, the handler is unaware of what the dog is doing and if it has found a casualty.

### 2.2 Urban Search and Rescue Challenges

Wireless network challenges include the materials found in a disaster area. The debris varies from concrete rubble, sewer pipes, rebar, and even vehicles [11]. USAR challenges include dogs venturing out of sight of their respective handlers [9]. This occurs because dogs are quite capable of crawling and getting into tight spaces. With their agility, they can quickly climb up and over debris piles leaving their handlers behind to catch up. In all of these instances the handlers are unaware of what the canine is doing and if they have found a casualty; the handlers are unaware of the situation.

Situational awareness (SA) is the perception of the elements in the environment within a volume of time and space, the understanding of their meaning and the projection of their status in the near future [21]. Cameras affixed on canines have been used to 10 wirelessly transmit video feeds to handlers. These feeds provide situational awareness regarding the disaster area [9].

These tools are helpful but only offer situational awareness from the perspective of the canine's point of view. The purpose of this research is to provide situational awareness as to what the canine is doing. This would increase the situational awareness provided to the canine handler.

## 2.3 Challenges in Wireless Mesh Networks

Wireless networks experience many challenges that are not present with wired networks. Adverse environmental conditions add to the challenges that these networks face; challenges include the weather, temperature, humidity, and surrounding materials, such as materials known to cause interference (lead, steel, rebar, and concrete) [22]. Network interference also includes the increasing number of wireless enabled devices like cell phones, desktops, laptops, smart phones, etc, all with the capability of Bluetooth, GPS, Wi-Fi and access to cellular networks base stations [23-25]. Wireless networks are complicated with nearly every factor affecting their ability to perform at their optimal speeds, as listed above.

Most academic research on WMNs has been conducted through simulation. This is partly due to the limited resources and high costs associated with purchasing the equipment required to conduct such experiments. Other reasons include the scale of the experiments; it is far easier to simulate a WMN with over 100 nodes then it is to test such a grand scale network. This was one of the constraints that we faced with our experiments. The WMN experiments conducted could have been extended through the use of additional mesh routers; however, this was not possible due to the limited number of mesh routers that were available for testing. Simulations are conducted with synthetic traffic patterns and node placement. As a result they do not produce realistic results as could be expected if the WMN were actually deployed.

### 2.3.1 Propagation Delay

All of the materials and environmental conditions listed in section 2.3 significantly affect the propagation delay of a WMN. It is not only the materials themselves that affect propagation delay, but their dimensions and thickness play a part, each a factor increasing the propagation delay within a network [22]. Some materials refract wireless signals, while others prevent them from penetrating through [22]. Moreover, there is an inverse relationship between the number of hops and performance when it comes to propagation delay [26, 27]; these factors contribute to increasing the propagation delay.

To the best of our knowledge the majority of published research work in the area of measuring propagation delay was based on simulation experiments with synthetic traffic and placement of nodes [26-29]. There was a study conducted by Microsoft Corporation [30], where propagation delay was evaluated across a WMN, which was deployed in an office building and used real user network traffic. The experiment involved a 21 node multi-radio WMN testbed.

Different office mesh network designs were assessed for their impact on the performance of the network. This research concluded that the captured user traffic was substantially dissimilar to the synthetic traffic used in similar experiments conducted through simulations. The results showed an additional median propagation delay of 20 ms with each transmission across the WMN, compared to simulation results.

Our experiments deployed an actual WMN in a building closely resembling that of a partially collapsed building. The results achieved produced actual propagation delays expected for the different configurations tested and the scenarios they represent in a disaster environment.

### 2.3.2 Packet Delivery Ratio

Using TCP protocols to transmit the data across the WMN could cause packet loss due to the window size, which may become congested and full. When this occurs the PDR decreases as packets are lost [30, 31]. There are three indications of packet loss when using TCP. The first indication is a retransmission timeout (RTO) at the source. The second is the arrival of duplicate acknowledgements (ACKs) at the source. Finally, the third indication is the receipt of the Internet Control Message Protocol (ICMP) source quench message [31].

TCP measures the length of time for an ACK to return from the destination also known as the Round Trip Time (RTT). The protocol keeps track of the average of this delay and estimates the deviation of the delay based on these averages. This delay is then used to determine if congestion is likely to occur. The protocol deems it likely there is congestion when the RTT delay is greater than four times the deviation estimated. In this case TCP runs congestion avoidance, which increases the congestion window [32-35]. This is done to ensure that packets are not lost and that the PDR remains high.

### 2.3.3 Wireless TCP

The TCP protocol is widely used and is effective in transmitting data packets to its destination. When TCP is utilized over a wireless network experiences some performance issues. One issue pertains to the propagation delay across the network, which may be increased. Another performance issue relates to packet delivery ratios, which may decrease. This occurs as packets are lost in the transmission of data across the network [31]. This paper presented a survey of different TCP performance improvement schemes for wireless networks. It determined that wireless networks were not as reliable as wired networks. TCP assumes that any packet loss that occurs is the result of congestion. TCP handles congestion by invoking congestion control. This works well in wired networks, but in wireless networks this results in decreased performance. Decreased performance occurs due to the characteristics of wireless networks, where packets are lost as a result of random high bit error rates and intermittent connectivity, which is due to the mobility of nodes and this could introduce long periods of disconnection.

## 3. WIRELESS MESH NETWORK PERFORMANCE

### 3.1 Propagation Delay Algorithm

The propagation delay algorithm was devised to minimize errors and ambiguities between the two systems (source and destination) on either end of the network. The laptop runs Microsoft Windows XP with a timestamp function for developers. The CPE device used a tiny microcontroller with an eight MHz frequency clock. The microcontroller clock and the operating system clock could not be synchronized since they are independent. This limitation was overcome by obtaining timestamps from the laptop in order to calculate the network's propagation delay.

### 3.1.1 Canine Pose Estimation Device Algorithm

The CPE device was now required to wait to receive a start bit identifier. This was denoted as ~, in the algorithm. For each start bit received, a data string was transmitted to the laptop as shown in the pseudocode below.

Canine Pose Estimation Device Algorithm Version 2 Pseudocode
```
    Start
        While (1)
            Wait (STARTBIT received from client program)
            If (STARTBIT == '~')
                Set (AccAx, AccAy, AccBx and AccBy to read (SerialPort))
                Transmit ("*AccAx AccAy AccBx AccBy \n")
                Delay (50 milliseconds)
```

  End

The client program on the laptop sent a byte of data across the Wi-Fi network to the nearest mesh router. Immediately after the byte was transmitted the Windows operating system time stamp was taken; the transmitted signal byte time stamp was denoted as Tx in the pseudocode below. The byte was transmitted along the mesh network hopping from node to node until it reached its final destination, the CPE device. This byte signaled to the CPE device to start collecting and transmitting accelerometer data across the network.

### 3.1.2 Computing Propagation Delay and Packet Delivery Ratio

```
Start
    Loop (until user hits control C) //ends application
        Open (serialPort )
        Connect (CPEdevice)
        Wait (STARTBIT received from client program)
        If (STARTBIT = '~')
            Write (STARTBIT to CPEdevice)
            Tx = Get (Windows Time Stamp)
            RequestCount = RequestCount + 1
            AccAx, AccAy, AccBx and AccBy = read (buffer)
            Rx = Get (Windows Time Stamp)
            PD = (Rx – Tx) / 2
            Write to File (PD)
            ReceivedCount = ReceivedCount + 1
            PDR = (RequestCount / ReceivedCount)
            OutputToFile (PDR)
            Display (PD, PDR)
End
```

Each mesh router used a mesh protocol to transmit the data across the network. When the CPE device received this signal, the CPE algorithm acquired acceleration measurements from the sensors and transmitted this data across the network to the laptop; a 22 byte string of the canine pose data was sent every 50 ms. This was comprised of the acceleration readings from both axes' of each accelerometer. When the pose data is received on the laptop, the Windows time stamp is taken and denoted as Rx.

### 3.1.3 Laptop Algorithm's Propagation Delay Formula

The experiments were conducted using typical traffic that would be transmitted across the WMN by the CPE system. The received timestamp was taken at the end of obtaining the entire 22 bytes of data. The reason behind using the actual CPE device data instead of a single byte was to determine the delay in receiving true accelerometer data across a WMN in a disaster environment; this was done to determine if a delay would be significant enough to impede canine pose estimation in real-time.

In reality, even these measurements while realistic for CPE data, would need to be retaken, since CPE is part of a much larger canine data system that would also be transmitting data over the same network. However, the implication of this larger data stream was beyond the scope of this research.

Depending on the pose, a canine typically takes between two and four seconds to perform a pose as determined by our experiments. If the delay was more than two seconds across the wireless mesh network, the CPE system would not be able to determine canine pose in real-time. As discussed previously, there are many factors that can significantly affect the connectivity and transmission of the data across the network.

If the system was not able to determine the canine pose in real-time, the handler may be misled by the system and act in accordance with a previously indicated situation – one that had already passed. This would be a significant issue as dogs are very agile and may be many meters away from their original position just seconds later.

Propagation delay is the time taken to transmit between source and destination nodes in a network [37]. To determine the propagation delay we measure the time it takes to send data from the source to its destination, Tx, and also measure the time it takes to receive the data sent back from the destination to the source, Rx. The difference between the Rx and Tx gives us the propagation delay of the data traveling across the network twice. To determine the experimental propagation delay from source to destination only, the resulting value was divided by two.

### 3.2 Canine Pose Estimation System Transmission Delays

Transmission delays are determined by the bandwidth of the channel, size of the packets being transmitted and the software sending the data to be transmitted [37]. There are many factors that need to be taken into consideration that can contribute some delay to the transmission of data across the network. One important consideration is the transmission rate of the Wi-Fi network using the CPE system.

#### 3.2.1 Wi-Fi Transmission Rate

Wi-Fi is theoretically capable of running up to 11 Mbps on an 802.11b network and up to 54 Mbps on an 802.11g network; this transmission speed cannot be realized due to the hardware in the CPE device. The CPE device was equipped with an eight MHz crystal that the microcontroller used as a frequency clock. Using this crystal, the fastest transmission rate possible was 38,400 Bd, to achieve the lowest tolerable error rate of +/- 0.2% [36].

Each string of data consisted of four acceleration readings. This meant that each string was 22 bytes or 176 bits in length. The time to transfer this data was calculated by taking the number of bits being transmitted and divided by the transmission speed of 38,400 Bd. This translated to a transmission rate of 4.58 ms per string of data. This rate was consistent for each transmission of the canine pose data. The first CPE device prototype employed Bluetooth networking and achieved a transmission rate of 48 000 Bd.

#### 3.2.2 Transmission Rate of Mesh Routers

Another delay that was factored into the experiment was that introduced by the mesh routers being used. These routers introduced 2 ms on average per hop, from mesh router to mesh router. The maximum delay could be calculated assuming a strictly linear sequence of mesh routers; this would have resulted in a delay of 8 ms (4 routers times 2 ms per router). Although this was the average delay known for the hardware, it could only be used as an estimated projection due to the self-configurable nature of the network and based on the number of nodes deployed. There was no other added delay by the mesh routing algorithm.

The transmission delay inherent with the CPE system included the hardware being used, protocols and algorithms. The first interface is the CPE device and has a transmission rate of 4.58 ms. The CPE algorithm's delay made up the second interface, with a maximum of 50 ms,. The third interface was the Serial to Wi-Fi module, which added up to 19 ms of additional transmission delay. The fourth interface was the mesh routers that could add a maximum of 8 ms additional delay (depending on the configuration). Thus the maximum transmission delay inherent to the system was 81.58 ms.

### 3.3 Packet Delivery Ratio

Packet Delivery Ratio (PDR) was defined as the number of packets received at the destination divided by the number of packets sent by the source [37]. This was an important metric to assess the reliability of the network in the transmission of the data. These experiments determined the number of packets being sent from the CPE device across the network, and the number of packets received by the laptop.

In order to accurately determine the PDR, the number of requests for the 22 byte data string sent from the laptop was tallied; this was denoted as RequestCount in the pseudocode found below. The number of complete data sets actually received by the laptop from the CPE device was also tallied and was denoted as ReceivedCount in the PDR pseudocode. The PDR was then calculated as the number of data strings received at the destination divided by the number of request packets sent from the source.

## 4. EXPERIMENTAL RESULTS AND DISCUSSION

The CPE device transmitted data utilizing a WMN, which broadcasted the data. The data was transmitted over the mesh network hopping from one mesh router to another until it reached its destination (the laptop). This multi-hop data transmission can experience signal loss and/or delays. It was important to evaluate and analyze whether the delay was significant enough to affect urban search and rescue. For example, if the dog was behind a wall and could not be seen by its handler, it would be imperative to know if the CPE system result would be accurate and had transmitted reliable data in near real-time. In the presence of obstacles and debris, the signal strength deteriorates from interference from many sources, as shown in Figure 1.

The WMN experiments were conducted with two performance metrics in mind: propagation delay and packet delivery ratio. These metrics enabled the assessment of the performance of the proposed CPE system. Propagation delay determined the expected time the data would take to travel across the WMN, while the packet delivery ratio was evaluated to determine if there was significant packet (data) loss. These metrics were also used to analyze different network configurations (node placement).

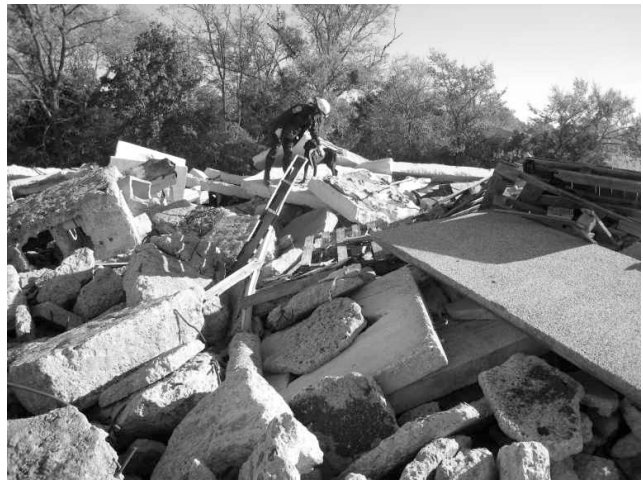

Figure 1. Handler rewards an USAR dog after the successful wireless activation of a drop bag. Testing indicated that the network's signal strength went from 100% to 0% 2.5 ft into any of the many holes that can be seen on the pile. Handler and dog TF1 (TX)

**4.1 Network Configuration and Experimental Setup**

The literature concerning WMN indicates that most previous work included testing propagation delay and packet delivery ratio through the use of simulators. The simulations were conducted using synthetically generated traffic. The traffic generally was comprised of TCP bulk transfers, which are typically transmitted randomly among the nodes in the WMN [26].

Experiments were conducted by deploying an actual WMN. The location of the deployment was essential in order to mimic that of an environment that would be found in USAR. The venue chosen the Center for Computing and Engineering (CCE, Ryerson University, Toronto) was a building with exposed concrete pillars and walls that would be similar to a USAR environment of a partially collapsed building. The building's structure was advantageous as all concrete walls and pillars were easily identifiable and could be used as barriers to simulate the environment found in a partial collapse scenario. The CPE device transmitted canine pose data across the WMN deployed in the building (Figure 2) in real time.

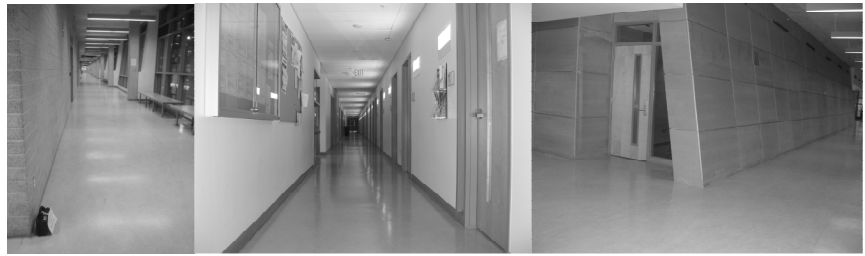

Figure 2. CCE Building of Experimental Environment (From left to right: Corridor 1, Corridor 2 and the Auditorium)

The WMN configurations as shown in Figures 5, 7, 9 and 11 were deployed in the building. It was ensured that each mesh router connected to the next mesh router, in order to meet the transmission of data across the network in correspondence with the configurations. The connection signal strength between each of the mesh routers was confirmed as a solid network connection with a signal to noise ratio, SNR below 60 dB and with signal strength no less than 70 dB. The last mesh router in the network acted like a gateway that connected the mesh network to the Internet. This mesh router was connected wirelessly to the Ryerson Network-Centric Applied Research Team (N-CART) lab's wireless network. The laptop connected to the Ryerson University wireless network. By setting up the network in this fashion, using two different network connections to the Internet, we ensure that the data received on the client end has successfully been transmitted from its destination point.

To confirm the network configuration, a test was performed to ensure that the data from the CPE device was being transmitted across the entire wireless mesh network and across the Internet, and received on the laptop. The mesh router was used as a gateway to the WMN with the Internet. Each configuration took approximately 2.5 hours to set up. Once all of the network nodes were connected, the distances and layout of the building were recorded. One hundred requests for Canine Pose data were transmitted from the laptop to the CPE device. This was repeated twice (listed as test 1 and 2) for each of the configurations.

This data was captured and written to an output file for later analysis. This building was chosen as the experimental environment due to the materials found in the structure of the building. Materials included steel reinforced concrete pillars and walls through out the building. Some of the other rooms in the building were made up of plain cinder block and/or wood and/or dry wall. It was assumed the walls would affect the WMN in a similar fashion as the rubble found in a partially collapsed building disaster scenario. The difference being was that the configuration of the rubble would be different from that found in this building.

**4.2 Wireless Mesh Network Configurations**

There were a few limitations with the use of this building. For one, only certain labs (1 through 6) and the hallway were accessible. The last mesh router in the network was restricted to lab 6 in order to access the NCART wireless network. Each of the configurations was comprised of four mesh routers and two wireless clients. The first node, mesh router 1 was connected to the CPE device and mesh router 2 as shown in Figure 3. All the mesh routers connected to the next mesh router. At the end of the WMN was mesh router 4, which was used as a gateway and connected to the Internet. The laptop also connected to the Internet and through that accessed the canine pose data.

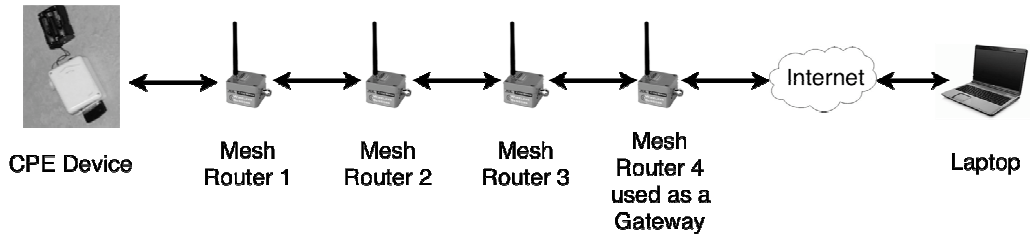

Figure 3. Configuration of Wireless Nodes Order

The building where the experiments were conducted was rectangular in shape with many rooms and corridors, which allowed for several different configurations and placement of the mesh routers. There were two straight corridors that ran North and South labeled hall 1 and 2 respectively in Figure 4. There were four corridors that ran east to West labeled as corridors 3, 4, 5 and 6.

From the results obtained from the conducted WMN experiments we look at two important network metrics, propagation delay and packet delivery ratios, for different WMN test-bed configurations. We compare the repeated tests and discuss the reliability of the results. The mean propagation delay was calculated for a data set, where a data set was comprised of ten canine pose data strings that were transmitted by the CPE device.

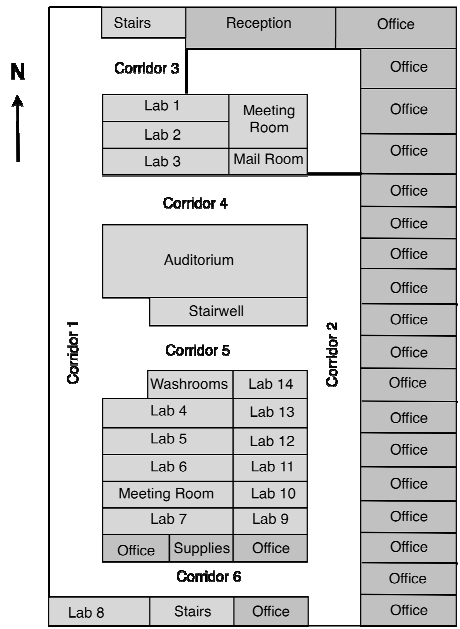

Figure 4. Building Layout

Comparing the measured propagation delay and PDR between each of the configurations we were able to determine if there were any significant differences, increases or decreases in the propagation delay and PDR relative to each configuration. We can then determine the best configurations that provide the lowest propagation delay and PDR, as well as the circumstances surrounding them.

## 4.3 Propagation Delay

This section analyzes the experimental data for each of the tests per configuration and assesses the consistency and repeatability of the tests. The propagation delay data was found to be normally distributed for each of the experiments conducted. It is important to note that there is an inherent transmission delay in the CPE System that would add a maximum additional 81.58 ms. This delay is a precursor to the propagation delay.

### 4.3.1 Configuration 1

The first configuration was a simple linear formation free from any obstacles as shown in Figure 5. This configuration represented the base case under optimal environmental conditions. The other configurations were compared to this first configuration in terms of propagation delay and PDR. The extended network coverage possible under such environmental conditions while maintaining good signal strength between each of the mesh nodes was also assessed.

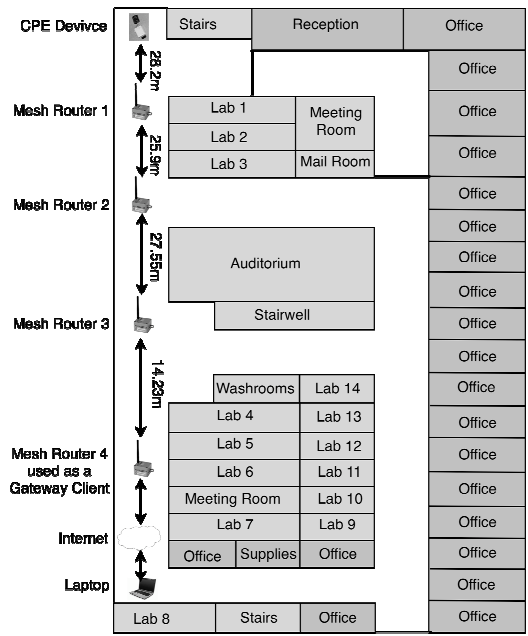

Figure 5. WMN Configuration 1

Configuration 1 was the baseline case to compare all the other configurations. This configuration was a measure of the best case scenario in the experimental environment as there were no impeding obstacles causing interference with the WMN. Network coverage was measured from the CPE device to the fourth mesh router. The physical distance from one end of the configuration to the other was 95.9 m. The first test had a mean propagation delay of 170.24 ms. The second test produced a mean of 318.42 ms. When comparing them with each other, there was a difference of 148.18 ms between the two means. Figure 6 shows the mean propagation delay for each data set. The mean propagation delay experienced by the WMN in configuration 1, was 244.33 ms.

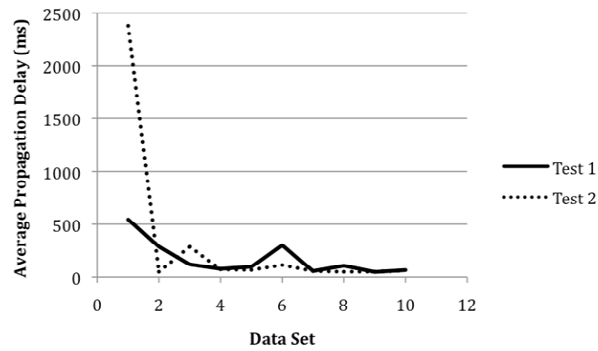

Figure 6. Mean Propagation Delay for Configuration 1

### 4.3.2 Configuration 2

The second WMN configuration is shown in Figure 7. and depicts extended network coverage. The mesh routers were deployed in a manner to extend network connectivity around obstacles that do not allow wireless radio signals to penetrate through. This configuration represents large thick obstacles made of reinforced concrete impeding wireless transmission and also where some rooms would be inaccessible and the USAR dogs would have to go around obstacles in order to continue searching.

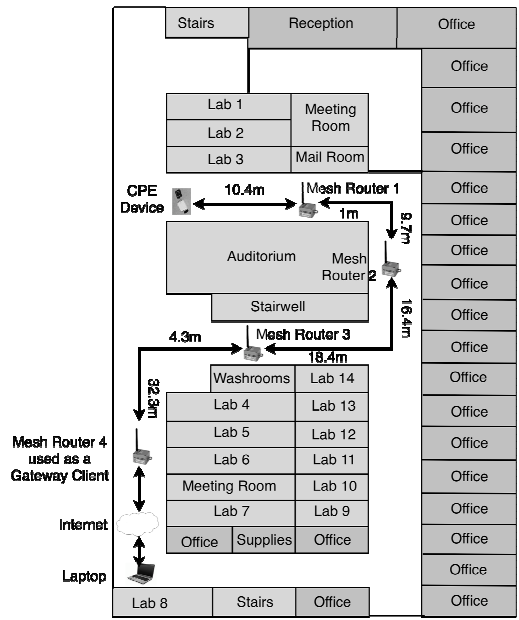

Figure 7. WMN Configuration 2

This configuration was chosen to test the ability of the WMN to extend network coverage around barriers that the signal cannot penetrate. This would have diminished the network's ability to transmit successfully and significantly increased the propagation delay. All of the nodes were placed behind obstacles as seen in Figure 8. The nodes could only connect to each other based on their transmission range.

This configurations network coverage was a total distance of 92.5 m. When compared to configuration 1, this configuration was 3.4 m shorter. In the first test, the mean propagation delay was 456.4 ms as shown in Figure 4.11. The second test had a mean propagation delay of 515.23 ms. This resulted in a difference of 58.83 ms between the configuration 2 results. The WMN had an mean propagation delay of 485.58 ms. There was a difference of 241.49 ms between this mean propagation delay and configuration 1 delay. This was a distinct measureable difference, showing that the propagation delay has significantly increased in this situation.

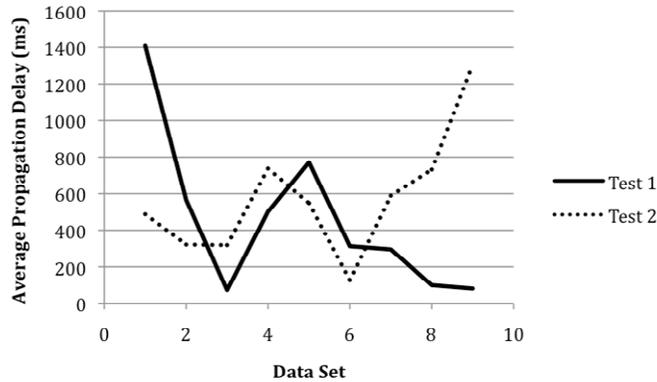

Figure 8. Mean Propagation Delay for Configuration 2

### 4.3.3 Configuration 3

The third configuration is shown in Figure 9 in where an attempt was made to penetrate through some obstacles by deploying a mesh router in one of the labs. In cases where a room has not caved in we may wish to extend the network into this room so that if a dog is searching in a large room the handler will still be able to receive the pose data at the other end of the network. Node placement is important to ensure that each of the nodes in the network is connected.

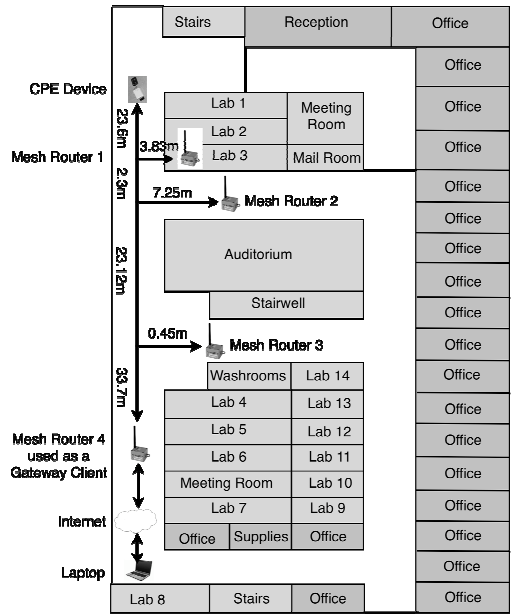

Figure 9. WMN Configuration 3

In configuration 3 there are obstacles placed directly between most of the nodes. Here we wish to determine the penetration power of the mesh routers. This is a great way to determine if a WMN could be deployed with nodes placed in different rooms. This would enable the dog to search an entire area without losing connectivity with its handler, when it moves from room to room, in a partially collapsed building.

For configuration 3 the first test had a mean propagation delay of 748.69 ms, shown in Figure 10. The second test has a mean of 664.5 ms. The difference between the two tests was 84.19 ms. In this configuration the WMN experienced an mean propagation time of 706.6 ms. This was a difference of 221.02 ms compared to configuration 2 and a difference of 462.51 ms compared to configuration 1.

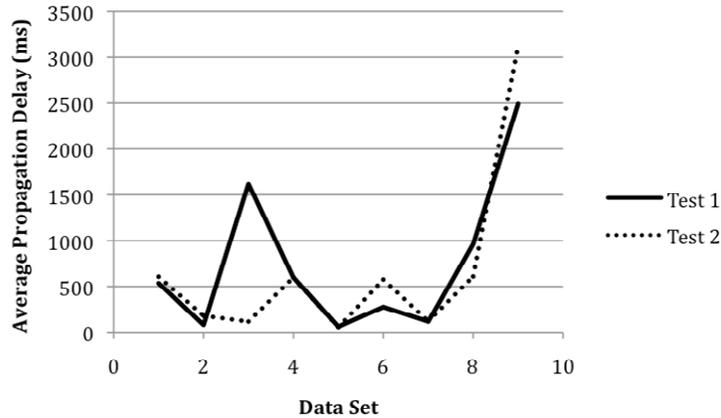

Figure 10. Mean Propagation Delay for Configuration 3

### 4.3.4 Configuration 4

The fourth WMN configuration is depicted in Figure 11. This configuration was attempted, but failed, because not all of the mesh nodes were unable to connect. A problematic node connection occurred between mesh router 1 and 2. Connectivity was established with less than 70 dB signal strength between nodes 2 and 3, and a slightly weaker connection was made between nodes 3 and 4. Modifications to the configuration were attempted without any success.

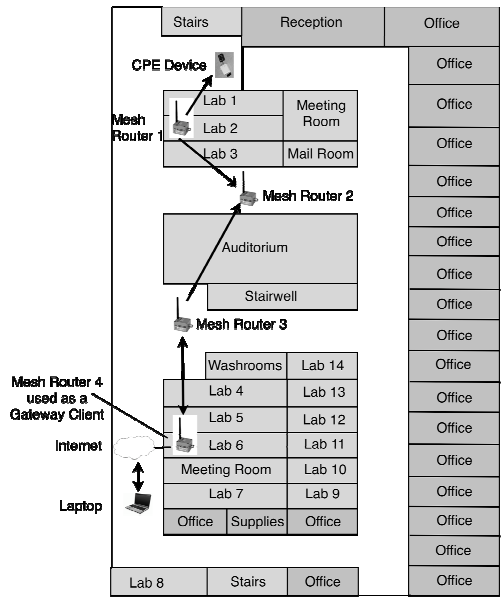

Figure 11. WMN Configuration 4

The following are the details of the attempted modifications. No connectivity could be achieved with router 1 in lab 1, lab 2, or lab 3. In addition, we tried router 1 just outside of each of these labs. In this case router 1 would connect to router 3 instead of router 2. Mesh router 1 was moved outside of lab 2 and router 2 was moved closer to router 1 sitting at a distance of 7.25 m from the width mid-point of corridor 1, along corridor 4. Router 1 could not be deployed further north in corridor 1 or it would lose connectivity with router 2. Router 2 dropped a few times causing more problems determining the placement of router 1. With router 2 any further into corridor 4 it would have lost connection with router 1 and 3. It was imperative that router 2 was close enough to router 1 to ensure good signal strength, so that router 1 would connect to router 2 and not directly to router 3.

This would occur if router 3 had a better signal than router 2, the mesh algorithm always chooses the best path. Even though router 3 was at a much greater distance, it was almost in direct line of sight thus having better signal strength despite the greater distance. As seen in Figure 4.14, router 3 is near the edge of the

Auditorium wall. Router 3 could not be deployed further east in corridor 5, otherwise it would lose connectivity with router 2. The farther east it was moved the lower the signal quality became between router 3 and 4. The signal strength between router 3 and 4 was weak, between 80 and 90 dB. In order to improve the connection between these nodes, router 4 was removed from lab 6 and placed in the corridor. This minimized the number of barriers the signal was required to penetrate to connect to the next node. The result was good connectivity. The signal strength decreased to 60 dB and a connection was established between the two nodes; however, the rest of the nodes were not able to establish a connection.

### 4.3 Packet Delivery Ratio

All of the tests in these experiments requested 100 data strings of the canine pose data. The PDR algorithm found a PDR of 100% for both tests for configuration 1 as seen in Figure 12. In configuration 2 there was a PDR of 88%, for both tests. While configuration 3 had a PDR of 84% for test 1 and 83% for test 2.

This shows that configuration 1, which experienced the least interference, was also the most reliable as it received all 100 packets at the other end of the network. This was a 100% PDR, with no packets lost. Configuration 2 was not as quite as successful receiving only 88% of its packets. This configuration experienced a higher degree of interference due to the node placement. Finally, configuration 3 had the lowest PDR of the configurations. It faced the most challenging environmental conditions with many obstacles directly between the nodes. There was a direct relationship between a high signal strength and PDR. The greater the signal strength, the greater was the PDR as a result (and vice-versa).

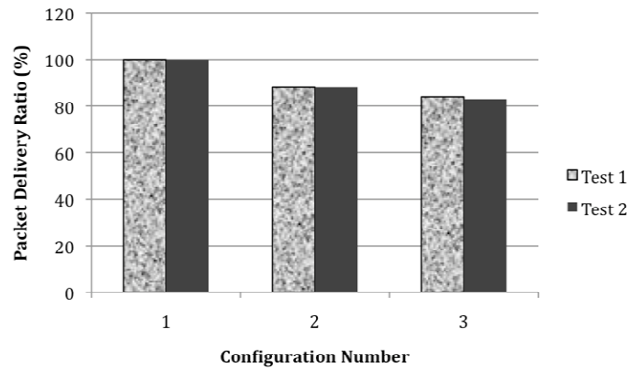

Figure 12. Packet Delivery Ratio for Configurations

### 4.4 Network Coverage

Comparison of the results found between each of the configurations provides insight into the importance of node placement as well as the impact obstacles have on the network. The performance of the network coverage was also assessed for each of the configurations using the signal strength and signal to noise ratio (SNR) metrics. Signal strength is counter intuitive, the higher the value the lower the performance. The lower the signal strength value the better was the established connection. A value higher than 70 dB was considered a poor connection and the network would drop intermittently or not connect at all. For the SNR any value below 60 dB made for a good connection between network nodes, representing low noise in the network.

#### 4.4.1 Network Coverage: Configuration 1

The network covered a distance of 95.9 m. A good connection was maintained with signal strength of 40 dB and a SNR of 20 dB. The signal strength was low and could be attributed to the fact that this configuration did not have any obstacles between the nodes. Although the SNR shows the network experienced some interference. In our experimental environment the causes of the interference that were experienced could include any of the following: the wireless network present thorough out the building, other wireless devices present in the building such as cell phones and smart phones, and finally the concrete and rebar in the buildings structure. This interference affected each of the configurations. This configuration portrays the optimal network connections possible in this testing environment, and was the base line for comparison of the other configurations.

#### 4.4.2 Network Coverage: Configuration 2

In this configuration the network connection was not optimal. The measured signal strength was 55 dB a difference of 15 dB when compared to configuration 1. The SNR was measured to be 42 dB, a difference of 22 dB from configuration 1. This added interference to the network is due to the nodes being placed out of the line of sight with the next node. As shown in Figure 4.10, each of the mesh routers were placed behind walls, in corridors 4, 2, 5 and 1, respectively. The close proximity to the walls made of concrete and steel rebar undoubtedly introduced more interference to the network.

#### 4.4.3 Network Coverage: Configuration 3

The physical distance that we were able to obtain network coverage for this configuration was 95.88 m linearly along corridor 1 with mesh routers 1, 2, and 3, deployed 3.83 m into lab 3, 7.25 m into corridor 4, and 0.45 m into corridor 5, respectively. This is deceptive, as to the real network coverage provided by the WMN. In configuration 1 as seen in Figure 4.8, the distance between each of the mesh routers are 25.9 m, 27.55 m and 14.23 m for a total of 67.68 m. When compared to configuration 3 in Figure 4.12, the distances are 2.3 m, 23.12 m and 33.7 m, totaling to 59.12 m.

Comparing the network coverage distance between these configurations, we found that the network coverage for configuration 1 was greater by 8.56 m. It is better to look at the direct distance between the nodes for this configuration to get a better idea of network coverage. When looking at the direct distance between each of the nodes, the network covers a distance of 88.34 m. This is much lower than that of the linear distance and indicates that barriers cause significant interference to the WMN. This interference, weakens the signal and diminishes the distance the signal can travel, thus the network coverage is decreased.

This was verified by the network performance metrics used to determine node placement. Signal strength was measured to be 67 dB; this was close to the tolerable threshold for a good connection, which was required to be less than 70 dB. The difference when compared to configuration 1 and configuration 2 was 27 dB and 12 dB respectively. The SNR was measured to be 56 dB, which is also near the tolerable threshold for a good connection (60 dB). A difference of configuration 3 compared to configuration 1 and configuration 2 was 36 dB and 14 dB respectively.

Configuration 3 had the highest levels of interference in the network, compared to the other scenarios. This was due to the nodes being placed in a room, or between rooms and having nodes connect to each other through the walls of varying materials. The network coverage of the first node with the second was a very short distance of 2.3 m apart and 3.42 m across; as compared to the distance between node 2 and 3 or 3 and 4 that were much greater. This was due to the difference in the thickness of the concrete walls.

## 5. CONCLUSION

The canine pose estimation (CPE) algorithm that was used to predict common poses of dogs in real-time. A wireless mesh network's (WMN's) ability to transmit canine pose in real-time in a disaster-like environment was assessed. The WMN's viability for use in USAR operations was also determined. We analyzed three metrics for wireless networks in USAR environments. These metrics included propagation delay, packet delivery ratio and network coverage. All experiments were conducted in what was assumed to be a disaster-like environment with real data. Other than the environment there was no simulation involved in testing the performance of the WMN with CPE data.

This research contributes a potential solution for providing additional situational awareness for USAR operations. Emergency first responders, search managers and canine handlers all stand to benefit from the use of the CPE system. It could contribute to decreasing search times and increasing the number of lives saved in urban disasters. In addition, a viable solution to the wireless network issues encountered in USAR environments was presented and its expected performance determined through measurement. The results of this research are presented below.

The CPE device had the ability to transmit data across a wireless mesh network (WMN) in the disaster environment. The reliability and speed of the network in the disaster zone was assessed by conducting experiments of propagation delay and packet delivery ratio (PDR) across the network.

The mean propagation delay was 244.33 ms, 485.58 ms and 706.6 ms for configuration 1, 2, and 3 respectively; all of which are well below this threshold, showing that canine pose data could successfully transmitted across a WMN in a disaster zone. The longest propagation delay measured across the network for configuration 3, was less than 3.25 times the minimum threshold and less than 5.25 times the maximum threshold. Showing that even in the worst case the propagation delay experienced was minimal and would not effect the transmission of canine pose data in real-time.

Obstacles introduced additional network interference causing PDR to decrease. In configuration 1, 2 and 3 the PDR was 100%, 88% and 83.5%, respectively. This showed that increasing the barriers and their proximity to the mesh nodes increased signal loss. The network coverage, or the distance between nodes obtained for the configurations was approximately 96 m for configuration 1, 93 m for configuration 2, and 96 m for configuration 3. This showed that increasing the number of obstacles significantly decreased network coverage. They also produced weak signal strength measurements, as well as decreased signal to noise ratios. These metrics were found to be the worst in configuration 3 and optimal in configuration 1.

For the purposes of USAR, a WMN was found to be a viable networking solution as was capable of penetrating walls (depending on their thickness) as was demonstrated in configuration 3. In cases where the mesh routers could not penetrate through an obstacle, a simple solution was to re-deploy closer to the obstacles. Another solution was to deploy the nodes around the obstacles and extend the network in this manner. This was shown in configuration 2; however, this method required more planning and may prove problematic during actual operations where expertise may not be available to optimize network routing performance.

The disaster environment's measured propagation delay did not significantly affect the results being obtained by the CPE data receivers. This was also true for the PDR and the WMN area coverage. Every disaster area is unique and thus the WMN requires field adjustments in order to obtain the optimal network conditions for data transmission. These adjustments should allow for successful deployment of the WMN and use of the CPE system under most USAR situations given the ability to place network nodes around the disaster site.

We hope this work contributes to improving canine search. Dogs are trained to search for people and they are very effective at it. Canines are fast and agile and will even let the handler know what they are doing or what they have found. This work is intended as a step in helping us understand the, often subtle, language of canine pose in order to help save lives.

## REFERENCES

[1] B. Wisner, B. Wisner, and Blaikie, P.M. "At Risk: Natural Hazards, People's Vulnerability and Disasters", Routledge, 2004.

[2] S. E. Chang, and N. Nojima, "Measuring post-disaster transportation system performance: the 1995 Kobe earthquake in comparative perspective," *Transportation Research Part A,* vol. 35, no. 6, pp. 475-494, 2001.

[3] P. Halpern, B. Rosen, S. Carasso *et al.*, "Intensive care in a field hospital in an urban disaster area: Lessons from the August 1999 earthquake in Turkey," *Critical Care Medicine,* vol. 31, no. 5, pp. 1410, 2003.

[4] C. R. Figley, R. E. Adams, S. Galea *et al.*, "Adverse Reactions Associated With Studying Persons Recently Exposed to Mass Urban Disaster," *The Journal of Nervous and Mental Disease,* vol. 192, no. 8, pp. 515, 2004.

[5] M. R. Endsley, "Situation awareness global assessment technique (SAGAT)", *Aerospace and Electronics Conference, NAECON 1988, Proceedings of the IEEE National*, vol. 3, no. 1, pp. 789-795.

[6] M. R. Endsley, "Toward a theory of situation awareness in dynamic systems", *Human Factors,* vol. 37, no. 1, pp. 32-64, 1995.

[7] M. R. Endsley, and DJ. Garland, "Theoretical underpinnings of situation awareness: A critical review", *Situation Awareness Analysis and Measurement*, Lawrence Erlbaum Associates, pp. 3-32, 2000.

[8] A. Ferworn, D. Ostrom, K. Barnum *et al.*, "Canine Remote Deployment System for Urban Search and Rescue", *Journal of Homeland Security and Emergency Management*. Volume 5, Article 9, 2008.

[9] A. Ferworn, D. Ostrom, A. Sadeghian *et al.*, "Canine as Robot in Directed Search", *IEEE International Conference on System of Systems Engineering,* San Antonio, USA, 2007, pp. 1-5.

[10] A. Ferworn, A. Sadeghian, K. Barnum *et al.*, "Urban Search and Rescue with Canine Augmentation Technology", *IEEE System of Systems Engineering (SoSE06)*, Los Angeles, USA, 2006.

[11] A. Ferworn, D. Ostrom, A. Sadeghian *et al.*, "Rubble Search with Canine Augmentation Technology", *IEEE International Conference on System of Systems Engineering*, San Antonio, USA. 2007, pp. 1-6.

[12] S. Watanabe, M. Izawa, A. Kato *et al.*, "A new technique for monitoring the detailed behaviour of terrestrial animals: A case study with the domestic cat", *Applied Animal Behaviour Science,* vol. 94, no. 1-2, pp. 117-131, 2005.

[13]  P. A. Kreutler, and D. M. Czajka-Narins, "Nutrition in perspective", Prentice-Hall Englewood Cliffs, NJ, 1980.

[14]  T. F. Sanquist, H. A. Mahy, C. Posse *et al.*, "Psychometric Survey Methods for Measuring Attitudes Toward Homeland Security Systems and Personal Privacy", Human Factors and Ergonomics Society, 2006.

[15]  M. Wolfe, "A study of police canine search teams as compared to officer search teams", *Canine training Articles,* T. U. S. P. C. Association (Editor), 1993.

[16]  K. G. Furton, and L. J. Myers, "The scientific foundation and efficacy of the use of canines as chemical detectors for explosives", *Talanta,* vol. 54, no. 3, pp. 487- 500, 2001.

[17]  I. Gazit, and J. Terkel, "Explosives detection by sniffer dogs following strenuous physical activity", *Applied Animal Behaviour Science,* vol. 81, no. 2, pp. 149-161, 2003.

[18]  J. M. Slabbert, and O. A. E. Rasa, "Observational learning of an acquired maternal behaviour pattern by working dog pups: an alternative training method", *Applied Animal Behaviour Science,* vol. 53, no. 4, pp. 309-316, 1997.

[19]  W. S. Helton, "Canine Models of Occupational Expertise", *Human Factors and Ergonomics Society Annual Meeting Proceedings, General Sessions*, pp. 875-879, 2006.

[20]  R. Fackrell, "Open area off leash search", *Canine training Articles*, U. S. P. C.Association (Editor), 1996.

[21]  M. R. Endsley, and D. J. Garland, "Theoretical underpinnings of situation awareness: A critical review", Lawrence Erlbaum Associates 2000.

[22]  A. Ferworn, N. Tran, J. Tran *et al.*, "WiFi repeater deployment for improved communication in confined-space urban disaster search", *IEEE System of Systems Engineering (SoSE07)*, 2007, pp. 1-5.

[23]  C. I. Shaw, R. M. Kacmarek, R. L. Hampton *et al.*, "Cellular phone interference with the operation of mechanical ventilators", *Critical Care Medicine,* vol. 32, no. 4, pp. 928, 2004.

[24]  S. Omar, and C. Rizos, "Incorporating GPS into Wireless Networks: Issues and Challenges", *The 6th International Symposium on Satellite Navigation Technology Including Mobile Positioning & Location Services*, Melbourne, Australia, 2003.

[25]  N. Golmie, R. E. Van Dyck, A. Soltanian *et al.*, "Interference Evaluation of Bluetooth and IEEE 802.11 b Systems", *Wireless Networks,* vol. 9, no. 3, pp. 201- 211, 2003.

[26]  D. Couto, D. Aguayo, J. Bicket *et al.*, "A high-throughput path metric for multihop wireless routing", *Wireless Networks,* vol. 11, no. 4, pp. 419-434, 2005.

[27]  J. Bicket, D. Aguayo, S. Biswas *et al.*, "Architecture and evaluation of an unplanned 802.11 b mesh network", *ACM New York*, NY, USA, 2005, pp. 31-42.

[28]  K. Farkas, O. Wellnitz, M. Dick *et al.*, "Real-time service provisioning for mobile and wireless networks", *Computer Communications,* vol. 29, no. 5, pp. 540-550, 2006.

[29]  P. Selvidge, "Examining tolerance for online delays", *Usability News,* vol. 5, no.1, pp. 1-5, 2003.

[30]  J. Eriksson, S. Agarwal, P. Bahl *et al.*, "Feasibility study of mesh networks for all wireless offices", *ACM/Usenix MobiSys,* 2006, pp. 69-82.

[31]  M. Patel, N. Tanna, P. Patel *et al.*, "TCP over Wireless Networks: Issues, Challenges and Survey of Solutions", *Technical report, Computer Science Department, University of Texas,* Dallas, pp. 1-22, 2001.

[32]  J. Postel, *"*Internet Protocol", *RFC 791*, September 1981.

[33]  J. Postel, "Transmission Control Protocol*"*, *RFC 791*, 1981.

[34]  W. R. Stevens, "TCP/IP Illustrated, Volume 1: The Protocols", Addison-Wesley Reading, MA, 1994.

[35] G. R. Wright, and W. R. Stevens, "TCP/IP illustrated. Vol. 2: The implementation", Addison-Wesley Professional Computing Series, Reading, Mass.: Addison-Wesley, 1995.

[36] Dimension Engineering. http://www.dimensionengineering.com/DEACCM2G.htm, 2008. (Last visited November 2008).

[37] M. Ma, M. K. Denko, and Y. Zhang, "Wireless Quality of Service: Techniques, Standards and Applications", Auerbach Publications, Taylor & Francis Group, USA, 2008.